# Prevalence of Small-scale Jets from the Networks of the Solar Transition Region and Chromosphere

H. Tian,[1]* E. E. DeLuca,[1] S. R. Cranmer,[1] B. De Pontieu,[2] H. Peter,[3] J. Martínez-Sykora,[2,4] L. Golub,[1] S. McKillop,[1] K. K. Reeves,[1] M. P. Miralles,[1] P. McCauley,[1] S. Saar,[1] P. Testa,[1] M. Weber,[1] N. Murphy,[1] J. Lemen,[2] A. Title,[2] P. Boerner,[2] N. Hurlburt,[2] T. D. Tarbell,[2] J. P. Wuelser,[2] L. Kleint,[2,4] C. Kankelborg,[5] S. Jaeggli,[5] M. Carlsson,[6] V. Hansteen,[6] S. W. McIntosh[7]

[1]Harvard-Smithsonian Center for Astrophysics, 60 Garden Street, Cambridge, MA 02138, USA.
[2]Lockheed Martin Solar and Astrophysics Laboratory, 3251 Hanover St., Org. A021S, Bldg. 252, Palo Alto, CA 94304, USA.
[3]Max Planck Institute for Solar System Research, Justus-von-Liebig-Weg 3, 37077 Göttingen, Germany.
[4]Bay Area Environmental Research Institute, 596 1st St West, Sonoma, CA 95476, USA.
[5]Department of Physics, Montana State University, Bozeman, P.O. Box 173840, Bozeman, MT 59717, USA.
[6]Institute of Theoretical Astrophysics, University of Oslo, P.O. Box 1029, Blindern, NO-0315 Oslo, Norway.
[7]High Altitude Observatory, National Center for Atmospheric Research, P.O. Box 3000, Boulder, CO 80307, USA.
*Correspondence to: hui.tian@cfa.harvard.edu

**Abstract**: As the interface between the Sun's photosphere and corona, the chromosphere and transition region play a key role in the formation and acceleration of the solar wind. Observations from the Interface Region Imaging Spectrograph reveal the prevalence of intermittent small-scale jets with speeds of 80-250 km s$^{-1}$ from the narrow bright network lanes of this interface region. These jets have lifetimes of 20-80 seconds and widths of ≤300 km. They originate from small-scale bright regions, often preceded by footpoint brightenings and accompanied by transverse waves with ~20 km s$^{-1}$ amplitudes. Many jets reach temperatures of at least ~10$^5$ K and constitute an important element of the transition region structures. They are likely an intermittent but persistent source of mass and energy for the solar wind.

**Main Text:**

The Sun continuously emits ionized particles into interplanetary space in the form of the solar wind. A challenging investigation has now carried on for almost 50 years to understand where the solar wind originates and how it is accelerated (*1, 2*). Dark regions in coronal images indicate the coronal holes that are the commonly accepted large-scale source regions of the high-speed solar wind. However, identifying precise origin sites within coronal holes requires high-resolution observations of the chromosphere and transition region (TR), a complex interface between the relatively cool photosphere (~6×10$^3$ K) and hot corona (10$^6$ K). The mass and energy that ultimately feeds the solar wind must pass through this region.

The dominant emission features in this interface region are the network structures that appear as narrow bright lanes enclosing dark cells, with sizes of ~20,000 km in radiance images of emission lines (*3*). The network lanes (networks thereafter) are believed to be locations of strong magnetic fluxes originating from the boundaries of convection cells with similar sizes in the photosphere. Previous observations of coronal holes with the Solar Ultraviolet Measurements of Emitted Radiation (SUMER) instrument (*4*) onboard the Solar and Heliospheric Observatory (SOHO) revealed Doppler blue shifts of 5-10 km s$^{-1}$ for emission lines formed in the upper TR (*5*). They were interpreted as signatures of the nascent solar wind guided by funnel-like magnetic structures originating from the networks (*6*).

Recent analyses revealed weak blue wing enhancements in profiles of emission lines formed in the TR (*7, 8*). These weak enhancements indicate the possible presence of a plasma component flowing upward with speeds of 50-100 km s$^{-1}$, which may provide heated mass to the solar wind (*8*). It has been difficult to test this proposed idea without direct imaging of such TR upflows on the solar disk. However, moderate-resolution observations have revealed signatures of chromospheric upflows being heated to TR temperatures at the solar limb in a coronal hole (*9*).

Using observations from the Interface Region Imaging Spectrograph (IRIS) (*10*), we report results from direct imaging on the solar disk of high-speed upflows with apparent speeds of 80-250 km s$^{-1}$. Thanks to the high resolution (~250 km) in new wavelength windows, IRIS slit-jaw imaging observations with the 1400Å, 1330Å, and 2796Å filters (see Supplementary Materials, SM thereafter) unambiguously reveal the prevalence of small-scale jet-like emission features from the bright networks (Figs. S1-S3, movies S1-S2). These three filters sample emission from the Si IV, C II and Mg II ions which are formed at temperatures of ~$10^5$ K, ~$3\times10^4$ K and ~$10^4$ K, respectively. These network jets usually show fast upward motion with no obvious downward component. Although these jets are more easily seen in coronal holes located near the solar limb (movies S1-S5), they are clearly detected at any location on the solar disk outside active regions (movie S6).

These network jets are best seen in 1330Å images. The jet widths are usually around ~300 km and approach the instrument resolution limit, suggesting that the actual widths of many jets may be even smaller. By applying the space-time technique (SM) to the 1330Å image sequence obtained on 23 January 2014 (Table S1, movie S2), we have quantified the apparent speeds and lifetimes for 63 randomly selected jets (Fig. 1). The speeds fall mostly in the range of 80-250 km s$^{-1}$, which is much larger than the sound speed and close to the Alfvén speed in the chromosphere (*11*) and TR. These velocities are significantly larger than previously reported jet velocities in the chromosphere and TR (*12-16*). Some jets also show signatures of acceleration. Their lifetimes range mainly from 20 to 80 seconds. Most jets extend to lengths of 4-10 Mm (1 Mm = $10^6$ m), although some clearly reach ~15 Mm.

Many network jets also exhibit obvious motions transverse to their propagation direction, indicating that they carry transverse magneto-hydrodynamic waves known as Alfvén waves (*11, 17*). The wave magnitudes are difficult to measure from slit-jaw images because strong emission from other features complicates the quantification of the transverse displacement, and the jet lifetimes are usually too short to allow the detection of a full wave cycle. Instead, we use spectroscopic observations to estimate the approximate velocity amplitudes of Alfvén waves. The root-mean-square value of the fluctuating Doppler shift of the Si IV 1393.77Å line is ~5 km s$^{-1}$, which can be regarded as the resolved wave amplitude (SM, Fig. S5).

Many of these network jets are likely the on-disk counterparts and TR manifestation of type-II spicules (SM), which are jet-like features moving upward with speeds of 50-110 km s$^{-1}$ in the chromosphere above the solar limb (*15, 16*). Our direct imaging of flows along these jets on the solar disk is almost unaffected by line-of-sight superposition, thus providing further support for the debated existence of high-speed jet-like features (*16, 18*). IRIS observations also reveal their origin in the networks, which off-limb observations cannot determine. Yet, we notice that network jet velocities are generally twice that of type-II spicules, suggesting that the network jets sampled by the TR passbands are those being heated and accelerated in the upper chromosphere and TR (*19*), and/or that the apparent speeds we observe here are not all caused by mass flows. Additional absorbing components at the blue wings of some chromospheric absorption lines were previously claimed to be on-disk counterparts of type-II spicules (*13*). These features with speeds of 20-50 km s$^{-1}$ are likely the lower-temperature parts and/or less-accelerated phase of the network jets.

Many network jets tend to recur at roughly the same locations on timescales of ~2-15 minutes. Our on-disk observations show that these jets originate from localized bright regions in the networks (Fig. 2, movie S4). Sometimes we see obvious brightening at the footpoints of these jets. A few jets appear to reveal the characteristic inverted "Y"-shape morphology (Fig. 2B) that is associated with a bipolar magnetic field line reconnecting with a unipolar large-scale field (*12*). These characteristics, together with the high speeds, suggest that some of these intermittent jets may result from repeated magnetic reconnection (*20*) between small magnetic loops and the background open flux in the networks. It is also possible that the source regions of these jets are too small to be resolved by IRIS, or that other mechanisms (SM) such as flux emergence and the associated Lorentz force are responsible for the acceleration of the jets (*21*).

Spectroscopic observations from IRIS reveal that many jets reach temperatures of at least ~10$^5$ K, the formation temperature of the Si IV 1393.77Å line under ionization equilibrium. The most prominent signature of network jets in Si IV line profiles is a significant increase of the line broadening, which could be a consequence of field-aligned flows (*22*) or unresolved transverse motions such as Alfvén waves (*23*) and twists (*24*). Combined imaging and spectral observations of IRIS can help evaluate the contribution from field-aligned flows and transverse motions.

Greatly enhanced widths of the Si IV line are found around two locations of network jets (Fig. 3). The slit crosses the lower part of a recurring jet complex at location 1. There the obvious enhancement of the line profile at the blue wing (Fig. 3D, SM) indicates an association with the network jets visible in the slit-jaw images (movie S5). Thus, the enhanced line broadening here is largely caused by the superposition of the field-aligned flows (jets) on the network background.

Location 2 corresponds to the upper part of some swaying network jets (movie S5). Given the nearly symmetric line profile and that this region is close to the limb, these jets are likely propagating largely in the plane perpendicular to the line-of-sight. So the line broadening appears to be largely caused by unresolved Alfvén waves, or small-scale twists which are often associated with unresolved torsional Alfvén waves (*25*). If we attribute the non-thermal width (SM, Fig. S5) to these unresolved waves, the wave amplitude is estimated to be ~21 km s$^{-1}$.

Intensity and line width maps of Si IV (Fig. 4) reveal details of the TR structures. One prominent feature of these maps is the presence of filamentary or elongated structures. Comparing these maps with the slit-jaw images (movie S6) reveals an association of many such features with network jets. Depending on viewing angles, enhanced line widths in these filamentary structures could be caused by either the superposition of jet emission on the network background, or unresolved transverse motions, or both. This association reveals that network jets constitute an important element of TR structures (SM).

These jets are likely an intermittent but continual source of mass and energy for the solar wind. We find a total mass loss rate of (2.8 - 36.4)×10$^{12}$ g s$^{-1}$ for these jets if we assume that all jet plasma contributes to the solar wind (SM). This value is about 2 - 24 times larger than the total mass loss rate of the solar wind, yet we have to remember that it is difficult to determine the true contribution to the solar wind without sufficiently sensitive coronal observations. With a wave amplitude of ~20 km s$^{-1}$, the energy flux of Alfvén waves carried by the jets should be 4 - 24 kW m$^{-2}$ (SM). This is much larger than that required to drive the solar wind (~700 W m$^{-2}$), yet we do not know how much of this energy is dissipated.

The prevalence of these network jets may challenge current solar wind models. Most time-steady descriptions of the solar wind (*1, 26*) rely on mass flux driven by evaporation from the upper TR, induced by a combination of downward heat conduction from the corona and local radiative losses (*27*). Although successfully predicting the coronal heating and wind properties at Earth, these models usually produce steady flows of only a few km s$^{-1}$ in the chromosphere and TR. Such steady low-speed outflows have never been imaged.

In contrast, our IRIS observations reveal the presence of intermittent high-speed upflows from the networks. If the mass in these jets actually is lost in the solar wind, then models must be updated to account for this highly intermittent component. A proposed reconnection-driven solar wind model (*6*) may be consistent with our observations. This scenario, which involves reconnection between open field lines in the network and surrounding low-lying loops, has been simulated numerically (*28*). However, the maximum outflow velocities produced by this model are only ~30 km s$^{-1}$, and it is unclear if the entire mass and energy flux of the wind can be produced in this way (*29*).

If these jets are not the nascent solar wind, at least their interaction with the wind should be considered in solar wind models since they are the most prominent TR features in the networks where the wind is believed to originate. One recent model does include some upward and downward motions of the TR plasma (*30*). However, these motions have speeds of ~60 km s$^{-1}$ at most, and the jets we observe show much faster upward motions. Obviously, a successful solar wind model must carefully evaluate the mass and energy contributions from these network jets.

**References and Notes:**

1. S. R. Cranmer, *Living Rev. Solar Phys.*, **6**, 3 (2009).
2. V. H. Hansteen, M. Velli, *Space Sci. Rev.* **172**, 89 (2012).
3. E. M. Reeves, *Sol. Phys.* **46**, 53 (1976).
4. K. Wilhelm et al., *Sol. Phys.* **162**, 189 (1995).
5. D. Hassler et al., *Science* **283**, 810 (1999).
6. C.-Y. Tu et al., *Science* **308**, 519 (2005).
7. B. De Pontieu et al., *Astrophys. J. Lett.* **701**, L1 (2009).
8. S. W. McIntosh et al., *Astrophys. J.* **727**, 7 (2011).
9. B. De Pontieu et al., *Science* **331**, 55 (2011).
10. B. De Pontieu et al., *Sol. Phys.* **289**, 2733 (2014).
11. B. De Pontieu et al., *Science* **318**, 1574 (2007).
12. K. Shibata et al., *Science* **318**, 1591 (2007).
13. L. Rouppe van der Voort et al., *Astrophys. J.* **705**, 272 (2009).
14. H.-S. Ji, W.-D. Cao, P. Goode, *Astrophys. J. Lett.* **750**, L25 (2012).
15. B. De Pontieu et al., *P. Astron. S. Japan* **59**, S655 (2007).
16. T. M. D. Pereira et al., *Astrophys. J.* **759**, 18 (2012).
17. S. W. McIntosh et al., *Nature* **475**, 477 (2011).
18. Y. Z. Zhang et al, *Astrophys. J.* **750**, 16 (2012).
19. J. Martínez-Sykora et al., *Astrophys. J.* **736**, 9 (2011).
20. V. Archontis et al., *Astron. Astrophys.* **512**, L2 (2010).
21. M. L. Goodman, *Astrophys. J.* **785**, 87 (2014).
22. B. De Pontieu, S. W. McIntosh, *Astrophys. J.* **722**, 1013 (2010).
23. J. Chae et al., *Astrophys. J.* **505**, 957 (1998).
24. B. De Pontieu et al., submitted to *Science* (2014).
25. A. A. van Ballegooijen et al., *Astrophys. J.* **736**, 3 (2011).
26. B. Chandran et al., *Astrophys. J.* **743**, 197 (2011).
27. R. Hammer, *Astrophys. J.* **259**, 767 (1982).
28. L.-P. Yang et al., *Astrophys. J.* **770**, 6 (2013).
29. S. R. Cranmer, A. A. van Ballegooijen, *Astrophys. J.* **720**, 824 (2010).
30. T. Matsumoto, T. K. Suzuki, *Astrophys. J.* **749**, 8 (2012).

31. J. R. Lemen et al., *Solar Phys.* **275**, 17 (2012).

32. W. Curdt et al. *Astron. Astrophys.* **375**, 591 (2001).

33. M. Carlsson, R. F. Stein, *Astrophys. J.* **481**, 500 (1997).

34. R. J. Rutten, B. De Pontieu, B. W. Lites, *ASP Conference Series* **183**, 383 (1999).

35. J. W. Cirtain et al., *Science* **318**, 1580 (2007).

36. J. Martínez-Sykora et al., *Astrophys. J.* **732**, 84 (2011).

37. H. Tian et al., *Astrophys. J.* **738**, 18 (2011).

38. J. T. Mariska, The Solar Transition Region, Cambridge: Cambridge Univ. Press (1992).

39. H. Peter, B. Gudiksen, Å. Nordlund, *Astrophys. J.* **638**, 1086 (2006).

40. E. Landi et al., *Astrophys. J.* **744**, 99 (2012).

41. S. Tomczyk et al., *Science* **317**, 1192 (2007).

42. D. Banerjee, D. Pérez-Suárez, J. G. Doyle, *Astron. Astrophys.* **501**, L15 (2009).

43. M. Hahn, E. Landi, D. W. Savin, *Astrophys. J.* **753**, 36 (2012).

44. T. J. Okamoto, B. De Pontieu, *Astrophys. J. Lett.* **736**, L24 (2011).

45. D. H. Sekse, L. Rouppe van der Voort, B. De Pontieu, *Astrophys. J.* **752**, 108 (2012).

46. T. van Doorsselaere et al., *Astrophys. J. Lett.* **676**, L73 (2008).

47. V. Yurchyshyn, V. Abramenko, P. Goode, *Astrophys. J.* **767**, 17 (2013).

48. S. W. McIntosh, B. De Pontieu, *Astrophys. J.* **707**, 524 (2009).

49. J. A. Klimchuk, *J. Geophy. Res.* **117**, A12102 (2012).

50. N. Nishizuka et al., *Astrophys. J. Lett.* **683**, L83 (2008).

51. M. J. Murray et al., *Astron. Astrophys.* **494**, 329 (2009).

52. J. A. McLaughlin et al., *Astrophys. J.* **749**, 30 (2012).

53. P. G. Judge et al. *Astrophys. J.* **746**, 158 (2012).

54. P. G. Judge, A. Tritschler, B. C. Low, *Astrophys. J. Lett.* **730**, L4 (2011).

55. U. Feldman, K. G. Widing, H. P. Warren, *Astrophys. J.* **522**, 1133 (1999).

56. H. P. Warren, A. R. Winebarger, *Astrophys. J. Lett.* **535**, L63 (2000).

57. A. H. Gabriel, *Philos. Trans. R. Soc. London A* **281**, 575 (1976).

58. J. F. Dowdy et al., *Solar Phys.* **105**, 35 (1986).

59. J. M. Beckers, *Solar Phys.* **3**, 367 (1968).

60. K. L. Harvey, F. Recely, *Solar Phys.* **211**, 31 (2002).

61. Y.-M. Wang, *ASP Conf. Ser.* **154**, 131 (1998).

62. T. M. D. Pereira et al., *Astrophys. J. Lett.* **792**, L15 (2014).

63. S. R. Cranmer, A. A. van Ballegooijen, *Astrophys. J. Supp.* **156**, 265 (2005).

64. G. L. Withbroe, R. W. Noyes, *Ann. Rev. Astron. Astrophys.* **15**, 363 (1977).

65. G. E. Brueckner, J.-D. F. Bartoe, *Astrophys. J.* **272**, 329 (1983).

**Acknowledgments:** IRIS is a NASA Small Explorer mission developed and operated by LMSAL with mission operations executed at NASA Ames Research center and major contributions to downlink communications funded by the Norwegian Space Center (NSC, Norway) through an ESA PRODEX contract. This work is supported by NASA contract NNG09FA40C (IRIS), the Lockheed Martin Independent Research Program, the European Research Council grant agreement No. 291058, contract 8100002705 from LMSAL to SAO, and NASA grant NNX11AO98G. The publicly available IRIS data (including the data used in this study) are archived at http://iris.lmsal.com/data.html.

References (31-64) are called out only in the Supplementary Materials (SM).

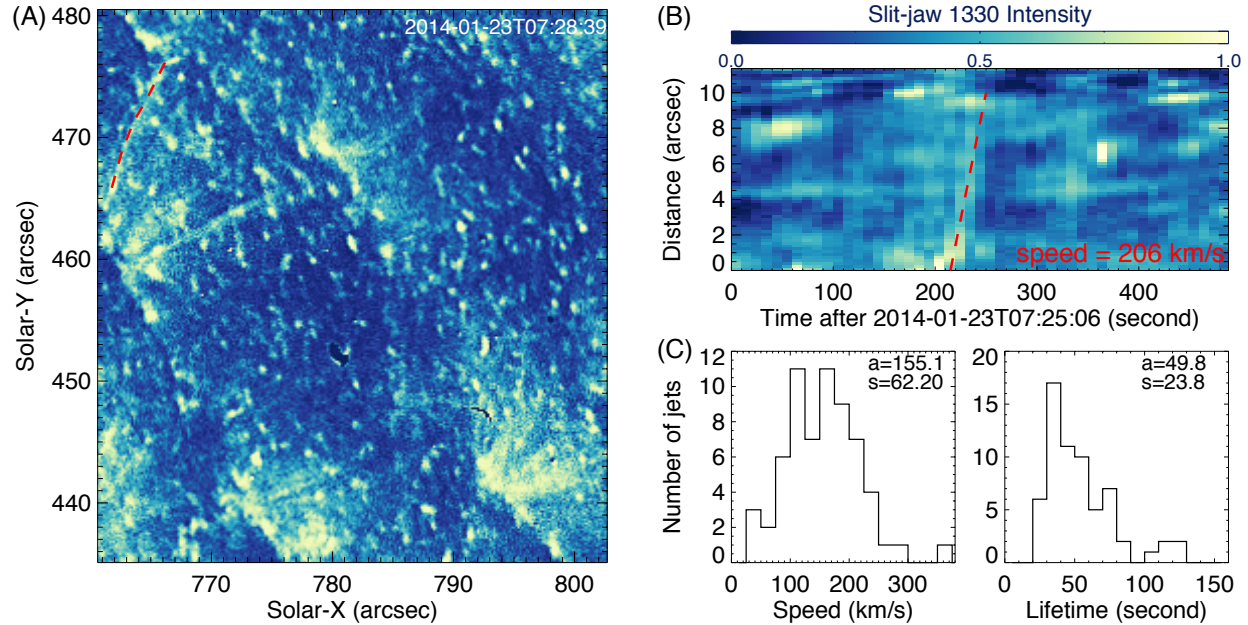

**Fig. 1**. Examples of network jets. (A) An unsharp masked (SM) 1330Å slit-jaw image (movie S3). The dashed line marks the path of a jet. (B) Space-time plot for the jet marked in (A). (C) Distributions of the apparent speeds and lifetimes for 63 jets. The average (a) and standard deviation (s) values are also shown.

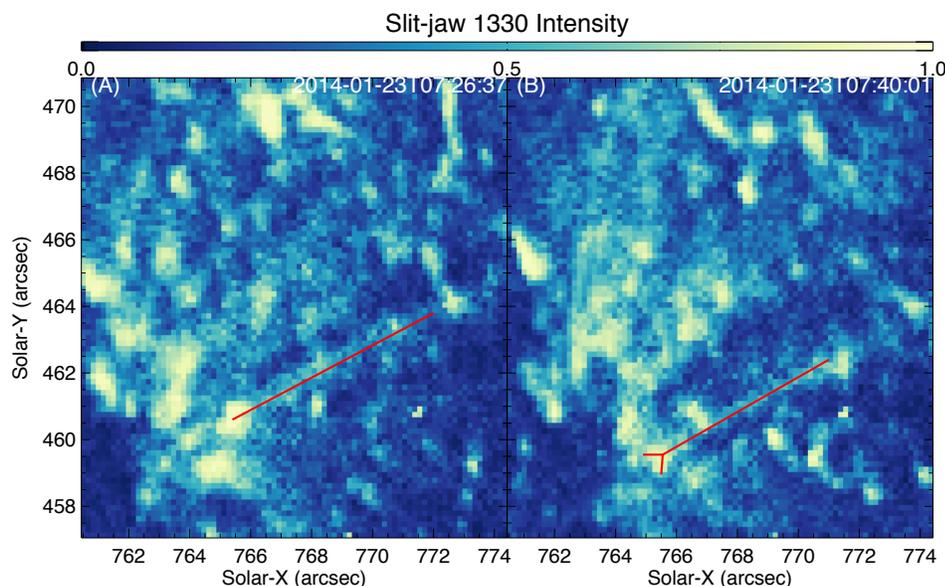

**Fig. 2.** Two unsharp masked 1330Å slit-jaw images showing the origin of network jets from small-scale bright regions in the network (movie S4). The red lines outline two jets.

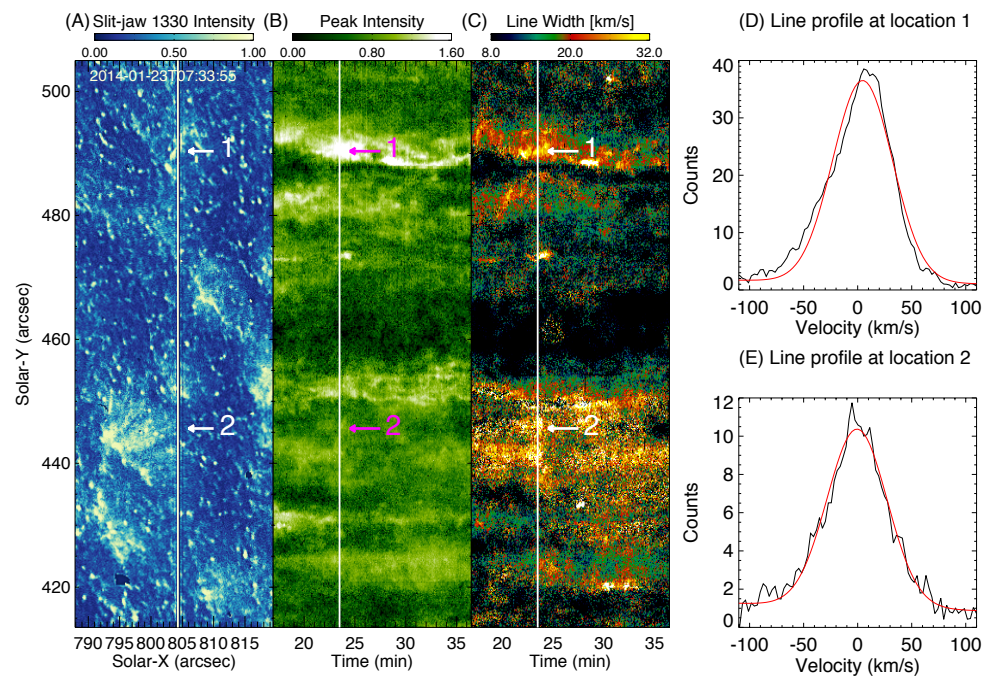

**Fig. 3.** Signatures of network jets in Si IV 1393.77Å line profiles (movie S5). (A) Unsharp masked 1330Å slit-jaw image taken at 07:33:55 UT on 23 January 2014. (B-C) Temporal evolution of the intensity and line width along the slit from Gaussian fit of Si IV line profiles. The vertical line indicates the slit location in (A) and time of 07:33:55 UT in (B-C). (D-E) Observed line profiles (black) at the two locations indicated by the arrows. Red lines are the Gaussian fits.

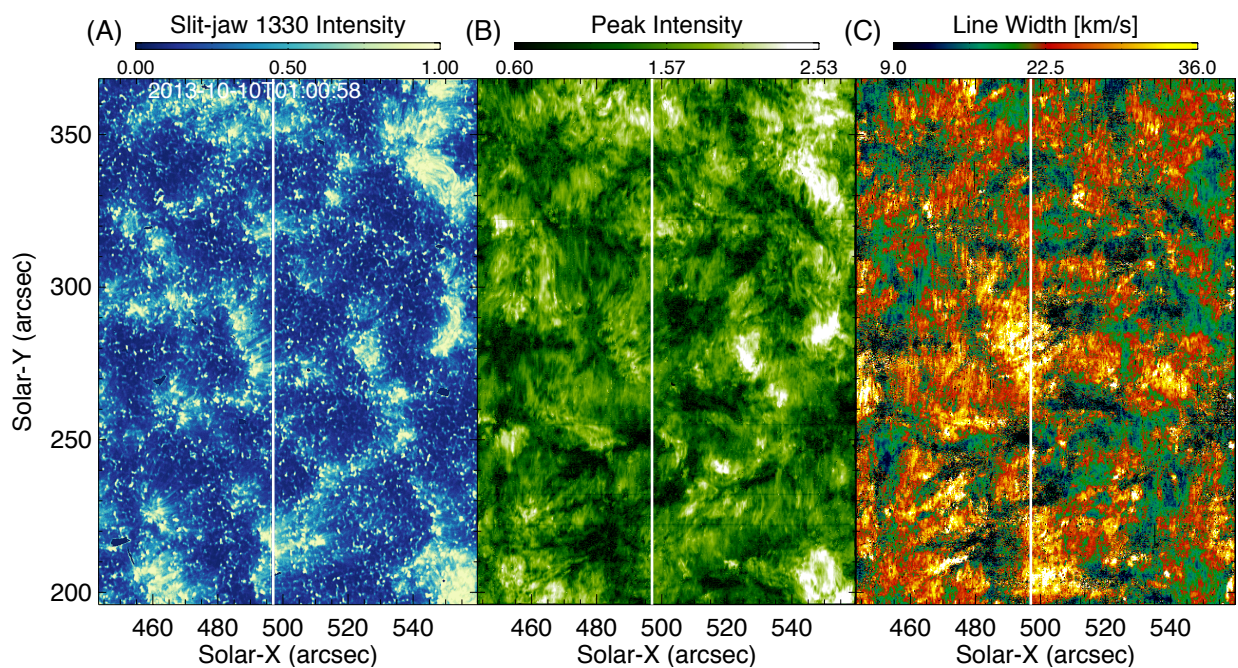

**Fig. 4.** TR filamentary structures caused by network jets (movie S6). (A) An unsharp masked 1330Å slit-jaw image. (B-C) Maps of intensity and line width from Gaussian fit of Si IV 1393.77Å line profiles. The vertical line indicates the slit location.

# Supplementary Materials for

# Prevalence of Small-scale Jets from the Networks of the Solar Transition Region and Chromosphere

H. Tian, E. E. DeLuca, S. R. Cranmer, B. De Pontieu, H. Peter, J. Martínez-Sykora, L. Golub, S. McKillop, K. K. Reeves, M. P. Miralles, P. McCauley, S. Saar, P. Testa, M. Weber, N. Murphy, J. Lemen, A. Title, P. Boerner, N. Hurlburt, T. D. Tarbell, J. P. Wuelser, L. Kleint, C. Kankelborg, S. Jaeggli, M. Carlsson, V. Hansteen, S. W. McIntosh

correspondence to: hui.tian@cfa.harvard.edu

**This PDF file includes:**

Supplementary Text S1 to S11
Figs. S1 to S5
Table S1
Captions for Movies S1 to S6

**Other Supplementary Materials for this manuscript includes the following:**

Movies S1 to S6

(http://www.sciencemag.org/content/346/6207/1255711/suppl/DC1)

**Supplementary Text**

**S1. Details of the observations**

Three data sets obtained with IRIS are used in this paper: a low-cadence sit-and-stare observation from 15:56 UT to 17:36 UT on 18 August 2013, a high-cadence sit-and-stare observation made from 07:10 UT to 08:02 UT on 23 January 2014, and a low-cadence raster scan from 23:26 UT on 9 October to 02:57 UT on 10 October 2013. Information on the pointing, size of the field of view (FOV) and exposure time is summarized in Table S1. The observed regions on the solar disk are shown as the rectangles outlined in the coronal images taken in the 211Å passband of the Atmospheric Imaging Assembly (AIA)(*31*) onboard the Solar Dynamics Observatory (SDO) in Fig. S1. Note that the observed region in the 23 January 2014 observation is part of a large coronal hole, although the coronal intensity seems not very weak due to the limb brightening effect and the foreground coronal emission from the surrounding quiet Sun. To demonstrate this we also show an AIA image taken one day before this observation, and mark the IRIS FOV and slit location on this image (Fig.S1A).

The 23 January 2014 observation was taken during the eclipse season of IRIS and we can see that the slit location drifted slightly to the east side. This leads to the fact that the signatures of the recurring jet complex shown at location 1 of Fig.3 move a little bit to the south on the slit since lower parts of the jets are crossed by the slit at later times.

The calibrated level 2 data of IRIS is used in our study. Dark current subtraction, flat field correction, and geometrical correction have all been taken into account in the level 2 data (*10*). The high-cadence observation (dataset 2) is used to study the network jet dynamics, e.g., lifetime, speed, moving direction. The low-cadence observations (datasets 1 and 3) cannot be used to study the dynamics of the jets since the time resolution is too low to fully resolve the propagation of the short-lived jets. But they are useful for the estimation of the spatial filling factor, comparison of the jet morphology in different passbands and study of spectral signatures of the network jets. Note that IRIS does not allow multiple passbands to be used at the same time. To achieve a high cadence, we often use only the 1330Å passband in our observations of network jets. Using multiple passbands alternatively in one observation will lead to a much lower cadence for each passband, which is often not enough for the study of the jet dynamics.

In this paper we use slit-jaw images mainly in the 1330Å passband, which samples emission from the strong C II 1334/1335Å lines formed in the lower TR ($\sim 3\times 10^4$ K). The network jets are also clearly present in the 1400Å passband, which samples emission from the Si IV 1394/1403Å lines formed in the TR ($\sim 10^5$ K). The contrast between the network jets and the background is lower in 1400Å images (Fig.S2, movie S1), which might be explained by the fact that in coronal holes the C II lines are about five times stronger than the Si IV lines (*32*). The broad passbands also include UV continuum emission formed in the upper photosphere. The ubiquitous grain-like structures in the slit-jaw images are mostly emission features of the UV continuum associated with granules (magneto-convective cells on the solar surface) and magneto-acoustic shocks (*33, 34*), and they usually do not have emission in the Si IV 1394Å line (e.g., Fig.4). The elongated jet-like features we report here can also be identified in the clean spectra of Si IV lines (e.g., Fig.4), thus are clearly TR features. The 2796Å passband samples emission from the Mg II 2796Å line formed in the upper chromosphere (~$10^4$ K). The contrast between the network jets and the background is even lower in 2796Å images (Fig.S2, movie S1), which might be caused by the lower resolution and the shadow of the Solc filter mask (in the upper right part of the images) for the 2796Å passband. But we can still see that many jets in this passband are clearly connected to the network jets in the TR passbands. The jets appear to be shorter in 2796Å images compared to the TR images.

## S2. Unsharp masking

Isolating individual network jets in the slit-jaw images is often not easy since the occurrence frequency of the jets is very high and different jets are often close to each other in space. The visibility of the jets is also hampered by the background network emission and emission from the ubiquitous bright grain-like structures in the network cells.

The unsharp masking technique we have applied to the slit-jaw images can make the network jet structures sharper in the images. When referring in the paper to an unsharp masked image we have subtracted a 1"×1" (6×6 pixels) box-car smoothed version of that image from the original image. The unsharp masked image is the total of this residual and the original image. To further enhance fine structures, we have also applied a Laplacian sharpening filter to these images. The visibility of the network jets in the unsharp masked image sequences is usually enhanced to a certain degree, although the quickly evolving grain-like structures are still present because of their similar spatial and time scales.

## S3. Space-time plots

The technique of space-time plots is widely used to derive the speeds of moving features (*35*). We choose the 23 January 2014 data set (movie S2) to study the speeds and lifetimes of the network jets. The observed region is large in size so that we can identify many network jets in this single observation. This observation also has a high cadence, which allows us to study the dynamics of the short-living jets. The observed region is also close to the limb, meaning that the line of sight component of the jet speeds should be small and that the jet velocities derived from the slit-jaw images should be close to the real velocities of the jets.

We have visually identified 63 jets with relatively strong emission in the 1330Å image sequence. These jets were observed at different times and they are well isolated from others, less affected by the bright grain-like structures and showing clear signatures in the space-time plots (see below). For each jet, we first draw a line (curved or straight) along the propagation direction (see an example in Fig.1A), then plot the intensity along this line and stack the intensity with time (Fig.1B). The lifetime and maximum length of the jet can be obtained directly from the space-time map. The minimum lifetime we derived is 20 seconds, which is limited by the cadence 10 seconds. It is likely that many jets have lifetimes shorter than 20 seconds. The speed can be calculated as the slope of the inclined stripe in the space-time plot. For example, the speed of a jet occurring around 07:28:39, marked in Fig.1A-B, is calculated as 206 km s$^{-1}$.

## S4. Asymmetry of line profiles

The technique of Red-blue (RB) asymmetry analysis was first introduced in 2009 to quantify the magnitude and velocity of the secondary emission component in some asymmetric line profiles (*7*). It is based on a comparison of the two wings of the line profile in the same velocity ranges. The blue wing emission integrated over a narrow spectral range is subtracted from that over the same range in the red wing. The range of integration will then be sequentially stepped outward from the line centroid to build an RB asymmetry profile (simply RB profile). The magnitude and velocity of the secondary emission component can then be inferred from this RB profile. Initially, the line centroid which separates the blue and red wings was simply taken as the centroid derived from a single Gaussian fit to the line profile (*7, 36*). Later on, it was found that in many situations using the spectral position corresponding to the peak intensity as the centroid can resolve the secondary component more accurately (*37*).

Here we apply the RB asymmetry technique to the line profiles shown in Fig.3D-E. These profiles have been averaged over 9 pixels along the slit to make the profiles smoother. We first apply a single Gaussian fit to the line profiles. However, the single Gaussian fit significantly deviates from the highly asymmetric line profile measured at location 1 (Fig.S4A). To reduce the large deviation of the Gaussian center from the position of the measured peak, we only fit the central 18 pixels around the peak with a single Gaussian. This Gaussian center is then taken as the centroid of the line profile. The spectral (velocity) bin size is chosen to be 10 km s$^{-1}$. The calculated RB profile is normalized to the peak intensity of the line profile. The RB profile (Fig.S4B) shows very large negative values at velocities of 20-70 km s$^{-1}$, suggesting significant contribution from upflows in this velocity range. Note that these velocities are the velocities of the upflows projected in the line of sight direction. The real velocities can be much larger. The average RB value in the velocity range of 20-70 km s$^{-1}$ is -0.31, indicating that the upflow intensity is about 31% of the background network intensity.

The line profile at location 2 (Fig.S4C) shows no strong enhancement at either wing, suggesting that the propagating directions of the associated network jets visible in movie S5 are largely perpendicular to the line of sight direction. Single Gaussian fits to the full line profile and the central part yield almost the same value of the line center. The RB profile (Fig.S4D) indicates only very weak blue wing enhancement and the average RB value in the velocity range of 40-130 km s$^{-1}$ is about -0.04, which is almost negligible.

## S5. Calculation of the non-thermal line width

The width of an emission line profile has contribution from both thermal and non-thermal motions. In addition, the instrument will also cause additional broadening of the line profile. So the observed line width can be expressed as (*38*):

$$\omega = \sqrt{\frac{2kT}{m} + \xi^2 + \sigma_I^2}$$

where $k$, $m$ and $T$ are the Boltzmann constant ($1.3806503\times10^{-23}$ m$^2$ kg s$^{-2}$ K$^{-1}$), mass and kinetic temperature of the emitting ion or atom, respectively. $\sqrt{2kT/m}$, $\xi$, $\sigma_I$ represent the thermal, non-thermal and instrumental width (all expressed in the velocity unit), respectively.

The calculation of thermal width depends on the determination of the kinetic temperature of the emitting ion or atom. The most common way is to use the formation temperature (temperature corresponding to the maximum of the contribution function) of the line as the kinetic temperature. Such a method is based on the assumption of (collisional) ionization equilibrium. The ionization and recombination times in transition region plasma are at least not (much) larger and often smaller than the typical hydrodynamic timescales. Thus, the assumption of ionization equilibrium usually can be justified (*39*). The formation temperature of the Si IV 1393.77Å line is about $10^{4.9}$ K (*40*). This will lead to a 1/$e$ width of the thermal broadening of 6.8 km s$^{-1}$.

According to the laboratory measurements before the launch of IRIS, the full width at half maximum of the instrumental profile is about 2.5 pixels for the wavelength band of Si IV 1393.77Å. Using the dispersion of 12.72 mÅ/pixel, the 1/$e$ width of the instrumental broadening expressed in the velocity unit can be calculated as 4.1 km s$^{-1}$.

For a measured total line width, the non-thermal component can then be easily calculated using the equation above.

## S6. Estimation of the Alfvén wave amplitude

As mentioned in the main text, the slit samples the upper part of a bunch of network jets around location 2 (Solar-Y=435" - 447", movie S5) shown in Fig. 3. These jets are recurring from the same network region many times during the whole observing period. These jets are largely lying in the plane perpendicular to the line of sight direction. So the Doppler shift fluctuation of the Si IV 1393.77Å line should give us information on the resolved component of the Alfvén wave amplitude (*41*). As we mentioned in the main text, the nearly symmetric line profile excludes significant contribution of field-aligned flows to the line broadening. Hence, small-scale transverse waves appear to be the major contributor to the non-thermal line width. Measurement of the non-thermal line width thus can be used to estimate the unresolved component of the Alfvén wave amplitude.

Fig. S5A-C shows the maps of the line parameters of the Si IV 1393.77Å line around slit location 2 marked in Fig.3. Based on the intensity and line width enhancement, and a comparison with the jets seen in the movie S5, pixels with intensity values (log counts) larger than 0.7 are regarded as samples of jets. If we interpret transverse motions associated with the jets as Alfvén waves, we can estimate the resolved and unresolved wave amplitudes from the measured values of Doppler shift and non-thermal width respectively at pixels within the contours.

The non-thermal line width is found to have an average value of ~21 km s$^{-1}$, which can be regarded as the unresolved Alfvén wave amplitude (root-mean-square value). This typical non-thermal width of our TR line is close to the previously reported values of non-thermal width at a temperature of ~$10^5$ K (*23*), and slightly smaller than the lower end of the non-thermal widths measured from coronal lines (formed above $10^6$ K) above the limb (*42, 43*).

The resolved Alfvén waves are associated with large-scale transverse motions, thus should show as a temporal fluctuation of the Doppler shift since the jets are nearly perpendicular to the line of sight at location 2 marked in Fig.3. We simply assume a zero average Doppler shift in the whole region since we are only interested in the fluctuation of the Doppler shift here. The root-mean-square value of the Doppler shift is found to be ~5 km s$^{-1}$, which may be regarded as the lower limit of the amplitude of the resolved Alfvén waves. This amplitude is only ~38% smaller than the amplitudes of the resolved Alfvén waves measured in even higher-resolution chromospheric (~$10^4$ K) images above limb (7.4 km s$^{-1}$) (*44*), and slightly smaller than the amplitudes of transverse motions associated with the chromospheric rapid blueward excursions (5-10 km s$^{-1}$) (*45*). This value is smaller than the amplitudes of low-frequency transverse waves in the chromosphere and corona (10-25 km s$^{-1}$, *11,17*).

The characteristic value of the total amplitudes of the transverse waves is then $\sqrt{21^2 + 5^2} \approx 22$ km s$^{-1}$. It is clear that the unresolved component (measured from the non-thermal width) dominates the wave amplitude. Interpreting the Doppler shift fluctuation as a consequence of kink waves (*46*), instead of resolved Alfvén waves (*41*), will change our estimation of the total Alfvén wave amplitude by only 1 km s$^{-1}$.

## S7. Comparison with previously reported jet-like features in the chromosphere and TR

The velocities of the network jets are generally 10 times larger than those of the chromospheric anemone jets outside sunspots (*12*) and impulsive ultrafine-scale ejections observed in active regions (*14*). These previously reported chromospheric (Ca II, He I lines) jets have velocities of only ~15 km s$^{-1}$, which were measured using the space-time technique.

Some network jets in the 1330Å images are likely the TR manifestation and on-disk counterparts of type-II spicules observed in the chromospheric Ca II images above the limb (*15, 16*). This suggestion may be supported by the following facts: (1) The linear jet morphology and transverse motions are similar for both features; (2) From movies S1 and S2 we see that the only elongated features on the disk and above limb are the network jets and spicule-like jets, respectively; (3) Movie S1 clearly reveals that at least some jets in the chromospheric and TR passbands originate from the same network regions. Yet, we noticed that the velocities of the network jets are generally twice larger than those of type-II spicules measured using the same space-time technique, suggesting an even larger contribution from these jets to the mass and energy of the upper atmosphere. We have noticed that the discovery of the network jets, regardless of their relationship to type-II spicules, is important since these intermittent high-speed jets originate from the solar wind source region and they contribute to the bulk TR emission.

Recently, rapid blueward excursions (RBEs) in the profiles of chromospheric absorption lines Hα and Ca II 8542Å were detected in several ground-based observations and they were claimed to be the on-disk counterparts of type-II spicules (*13, 45, 47*). The velocities of these additional absorbing features have been measured through both the space-time technique and spectral line profile asymmetries and they are usually in the range of 20-50 km s$^{-1}$, much smaller than the velocities of the network jets. It is likely that the RBEs are signatures of the lower-temperature and less-accelerated parts or phase of the network jets.

The extremely weak (2-5%) blueward asymmetries of SUMER line (e.g., C IV formed around $10^5$ K, Ne VIII formed formed around $6 \times 10^5$ K) profiles (*7-8*) seem to be related to the network jets. They both occur at the networks. The recurring frequency of the blueward asymmetries (*48*) is also similar to that of the network jets. It is likely that IRIS is now performing direct imaging of the high-speed upflows inferred from these profile asymmetries. However, we have noticed that the velocities of our network jets are generally 2-3 times larger than those inferred from the line profile asymmetries, which might be partly caused by different viewing angles and resolutions. The direct high-resolution imaging of IRIS has greatly reduced the uncertainty in the measurement of the jet parameters.

We noticed that several upward propagating jets with speeds up to 400 km s$^{-1}$ have been identified from the TR line profiles obtained by the HRTS rocket flights (*65*). A one-to-one correlation between these jets with spicules was excluded by these authors. Although the resolution of HRTS is ~1" and thus could not resolve the tiny jets we report here, we may not exclude the possibility that some of these high-speed spectral features result from the composite of several network jets.

IRIS observations have clearly demonstrated that the network jets can reach a temperature of ~$10^5$ K. Although we have found signatures of a few network jets in the 171Å passband of SDO/AIA in one observation we examined, it is still difficult to estimate how much of the jet plasma is heated to coronal temperatures since many of the network jets can not be resolved by the moderate-resolution AIA instrument. Moreover, the AIA 171Å passband includes contributions from not only coronal lines (Fe IX/X, ~$10^6$ K) but also TR lines (e.g., O IV, O V, ~$10^{5.2}$ K). Thus, tracing these jets to even higher temperatures and evaluating their contribution to coronal heating (*9, 49*) requires high-resolution observations of pure emission lines formed at higher temperatures. This should be one of the most important goals of future spectroscopic observations and high-resolution imaging in cool coronal lines, e.g., a Fe IX/X passband in a future Hi-C rocket flight.

## S8. Generation and acceleration mechanisms of the network jets

As we mentioned in the main text, the dynamics of the footpoints and the high speeds (close to the Alfvén speed in the interface region) suggest that recurrent magnetic reconnection between small-scale magnetic loops and the background network fields might be the mechanism for some network jets. The observed transverse waves accompanying these jets are also consistent with a numerical experiment of reconnection jets (*50*), where Alfvén waves are generated and propagating away from the reconnection regions. However, possible inverted "Y"-shape structures are found for only a few jets, suggesting that sizes of the source regions of these jets are below the resolution limit of IRIS or that there are other mechanisms responsible for the generation of the network jets.

Some numerical simulations have been performed to investigate the generation and acceleration mechanisms of small-scale jet-like features in the interface region. In a three-dimensional magneto-hydrodynamic (MHD) simulation, strong Lorentz force associated with large field gradients and intense electric currents squeezes the dense chromospheric material, resulting in a vertical pressure gradient that produce, heat and accelerate type-II spicules (*19*). Oscillatory reconnection between emerging fluxes and the coronal hole background fields has been demonstrated to be able to generate quasi-periodic upflows (*51, 52*). Reconnection between open fields in the chromospheric network and magnetic loops advected by supergranular flows has also been simulated and suggested to release mass to the solar wind (*28*). However, the speeds of the upflows generated in these models are mostly lower than 60 km s$^{-1}$. Our IRIS observations reveal much higher speeds at TR temperatures. The discrepancy between these existing models and our IRIS observations suggests the need of updating these models.

A recent numerical investigation has simulated Lorentz-force driven jets (*21*). This model has successfully produced speeds as high as 66-397 km s$^{-1}$. The different speeds depend on the different treatment of the viscosity. This study has demonstrated that magnetic forces must play an important role in the generation and acceleration of high-speed jets in the chromosphere and TR. Pressure driven jets produced by both the three-dimensional MHD simulation (*19*) and one-dimensional hydrodynamical models including time-dependent ionization (*53*) hardly reach speeds higher than 60 km s$^{-1}$.

There has been a suggestion that fine structures in the solar atmosphere such as spicules could be warps in two-dimensional sheet-like structures (*54*). This scenario may be consistent with the thin and elongated morphology of the bright network jets. However, this scenario has difficulty in explaining the apparent upward motion that is clearly observed for the network jets.

## S9. Relationship between elongated TR structures and network jets

Small-scale narrow loop like structures were previously reported in the intensity images of some TR lines observed with SUMER (*55, 56*). And it has been conjectured that some of these structures are related to spicules (*56*). This proposed idea could not be tested in the SOHO era due to the lack of TR imaging with sufficient sensitivity and resolution.

The spatial resolution of IRIS is about six times higher compared to SUMER. IRIS observations have revealed many more such filamentary structures in not only the intensity images, but also the images of line width. As we explained in the main text, the greatly enhanced line width is the primary signature of network jets in the Si IV line. Moreover, combined imaging and spectroscopic observations of IRIS allow us to establish a direct link between many of these filamentary structures and the highly dynamic network jets, as can be seen from Fig.4 and movie S6.

This connection has important implication for the TR structures. It is known that the observed TR emission in the temperature range of $10^4$ K - $2\times10^5$ K is much higher than that produced by models dominated by field-aligned heat conduction (*57*). This puzzle has led people to propose different scenarios of TR structures in the past decades, e.g., low-lying loops within the networks (*58*) and unresolved fine structures thermally disconnected from the corona (*55*). IRIS observations clearly reveal that many network jets reach at least ~$10^5$ K and that these jets must contribute to the emission of the TR significantly. Thus, these network jets should not be ignored when constructing TR models.

### S10. Mass loss rate of the network jets

The mass loss rate or mass flux of the network jets can be calculated as:

$$m = 4\pi r^2 \rho V f_t f_s f_h$$

where $r$ is the solar radius. Using an average lifetime of 50 seconds and a recurring period of 8.3 minutes, the temporal filling factor ($f_t$) of these jets is estimated to be ~0.1. A very rough estimation of the spatial filling factor ($f_s$) yields a value of ~1% from the 1330Å slit-jaw images. We use the typical measured number density ($N_e$) of spicules $(3.4 - 22)\times 10^{10}$ cm$^{-3}$ (*59*) since these jets are likely heated spicules (movie S1) which are fed with the dense chromospheric material (*19*). We assume a reasonable Helium abundance 5% and the mass density ($\rho$) can then be calculated as $1.1\times N_e \times m_p$, where $m_p$ is the proton mass. If we then take a typical fraction of the area of coronal holes ($f_h$) relative to the total area of the solar surface 5%-10% (*60*) and a measured jet speed ($V$) 150 km s$^{-1}$, we can obtain a mass loss rate of $(2.8 - 36.4)\times 10^{12}$ g s$^{-1}$ if all mass of the network jets is lost eventually.

The mass loss rate of the solar wind varies from ~$1.2\times 10^{12}$ g s$^{-1}$ in the solar minimum to $1.9\times 10^{12}$ g s$^{-1}$ in the solar maximum (*61*). The total mass loss rate of the network jets we estimated is comparable to or larger than the mass loss rate of the solar wind. However, lacking coronal observations of sufficient sensitivity it is difficult to determine the true contribution to the solar wind mass flux, especially since there are indications that some type-II spicule plasma may fall back down at TR temperatures (*9,62*). In our high-cadence (10 s) observation we do not see obvious downward motions. But it is also possible that emission from the downflowing plasma is too faint against the strong background to be seen by IRIS slit-jaw images. An examination of off-limb images of several coronal holes suggests possible signatures of downward motions following some jet-like upward motions. But the line-of-sight superposition effect makes it difficult to estimate how many jets show real downward motions since swaying motions of overlapping jets can easily lead to wrong identification of downward motions.

It might also be possible that some apparent motions are caused by thermal evolution or rapid ionization changes in a dynamic heating environment, or propagation of shocks. However, the presence of blue wing asymmetries in the Si IV lines suggest that at least a significant fraction of the apparent motions are clearly mass flows.

### S11. Energy flux of the Alfvén waves carried by the network jets

In many solar wind models (*1*), interaction between the upward and downward propagating Alfvén waves in the solar TR leads to Alfvénic turbulence which then accelerates the solar wind.

Since Alfvén waves are likely volume filling (*63*), we can assume a filling factor of unity for these waves. The energy flux associated with the Alfvén waves can be estimated as

$$f = \rho \langle \delta v^2 \rangle V_A$$

where $\rho$, $\delta v$ and $V_A$ are the mass density, velocity amplitude and Alfvén speed. The wave amplitude has been measured as ~20 km s$^{-1}$ from the Si IV line formed in the TR. As we mentioned in the main text, some of these jets are likely produced by magnetic reconnection. The measured jet speed is thus likely the lower limit of the Alfvén speed. If we use the measured jet speed 150 km s$^{-1}$ as the Alfvén speed and take the mass density of $(6.2 - 40.4) \times 10^{-14}$ g cm$^{-3}$ as calculated above, we can obtain an energy flux of 4 - 24 kW m$^{-2}$. The energy flux that reaches the corona will be enough to drive the solar wind (~700 W m$^{-2}$) (*64*) if the transmission coefficient is larger than ~3%. But it is still an open question how much of this energy is dissipated.

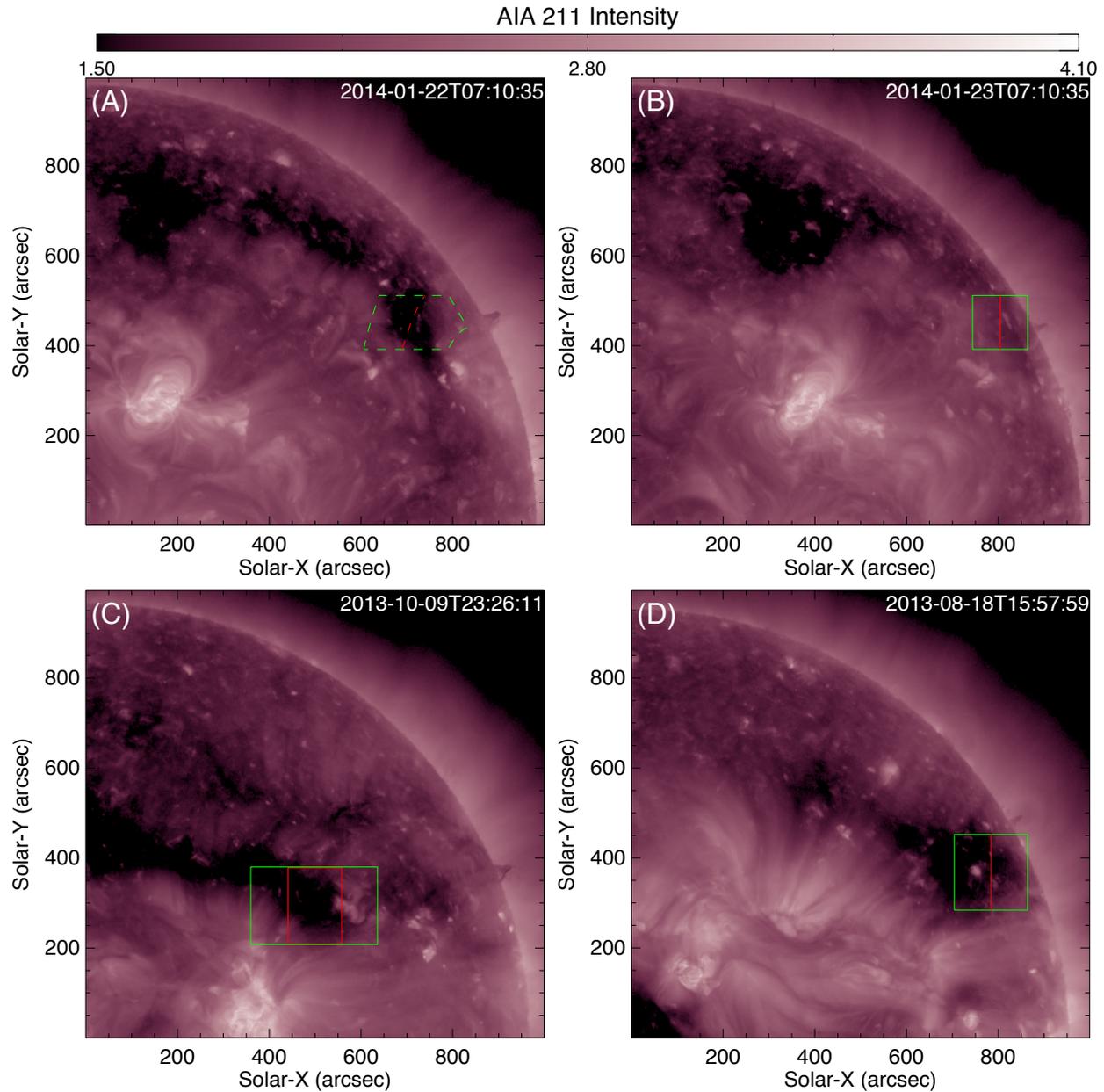

**Fig. S1. IRIS observation regions outlined in AIA 211Å images.**
(A)-(B) The 23 January 2014 observation. The green and red lines outline the FOVs of the slit-jaw imaging and spectroscopic observations, respectively (B). An AIA image taken one day before the 23 January 2014 observation is shown in (A). The dashed green and red lines represent the regions corresponding to the FOV and slit location in this observation, respectively. (C) The 9-10 October 2013 observation. (D) The 18 August 2013 observation. The intensities are in logarithmic scale.

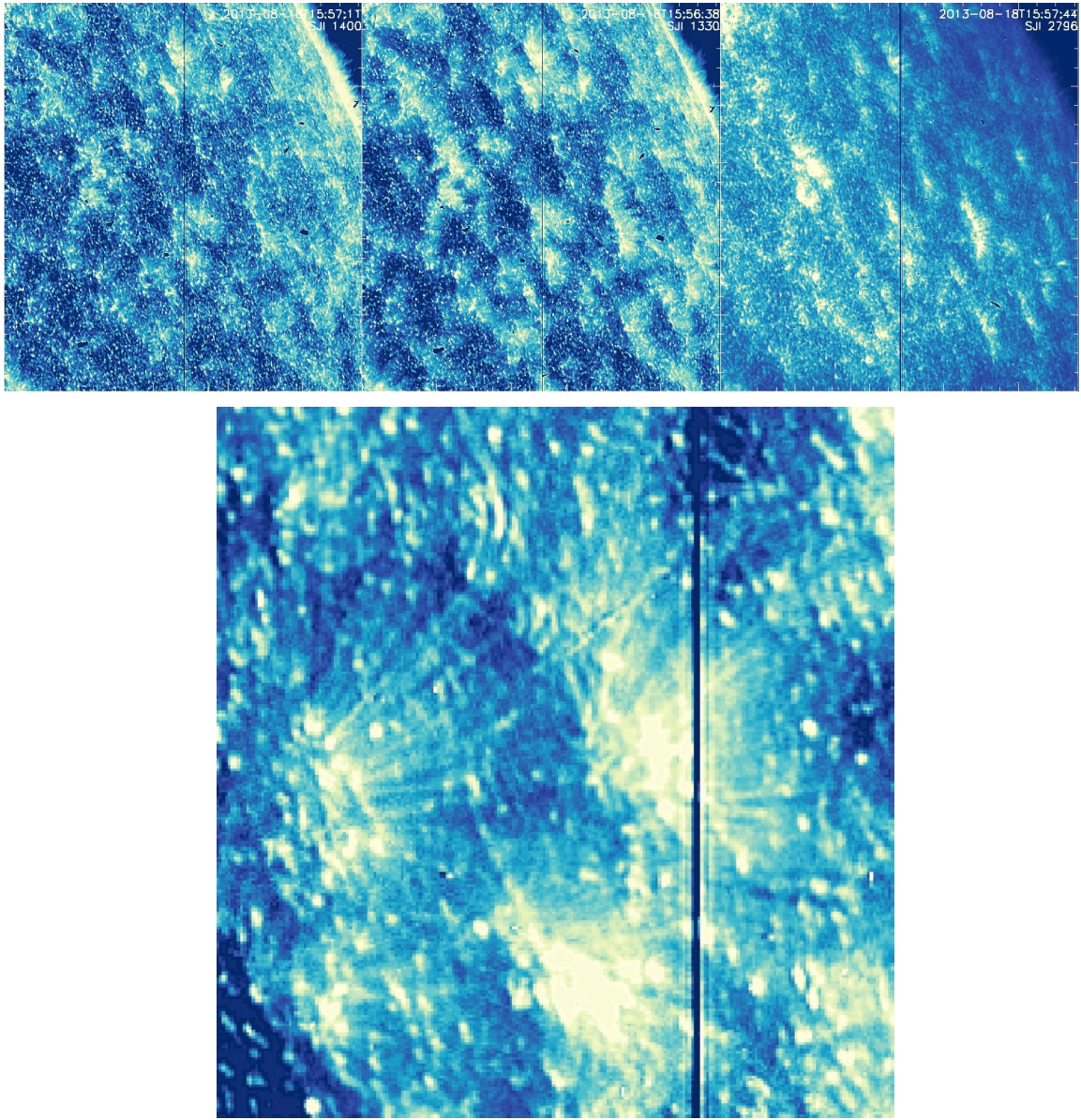

**Fig. S2. Network jets in different passbands.**
Upper: Unsharp masked 1400Å (left), 1330Å (middle) and 2796Å (right) slit-jaw images (SJI) obtained around 15:57 UT on 18 August 2013. The field of view is outlined in Fig. S1D. This is a snapshot of movie S1. Lower: A 1330Å image in a smaller FOV.

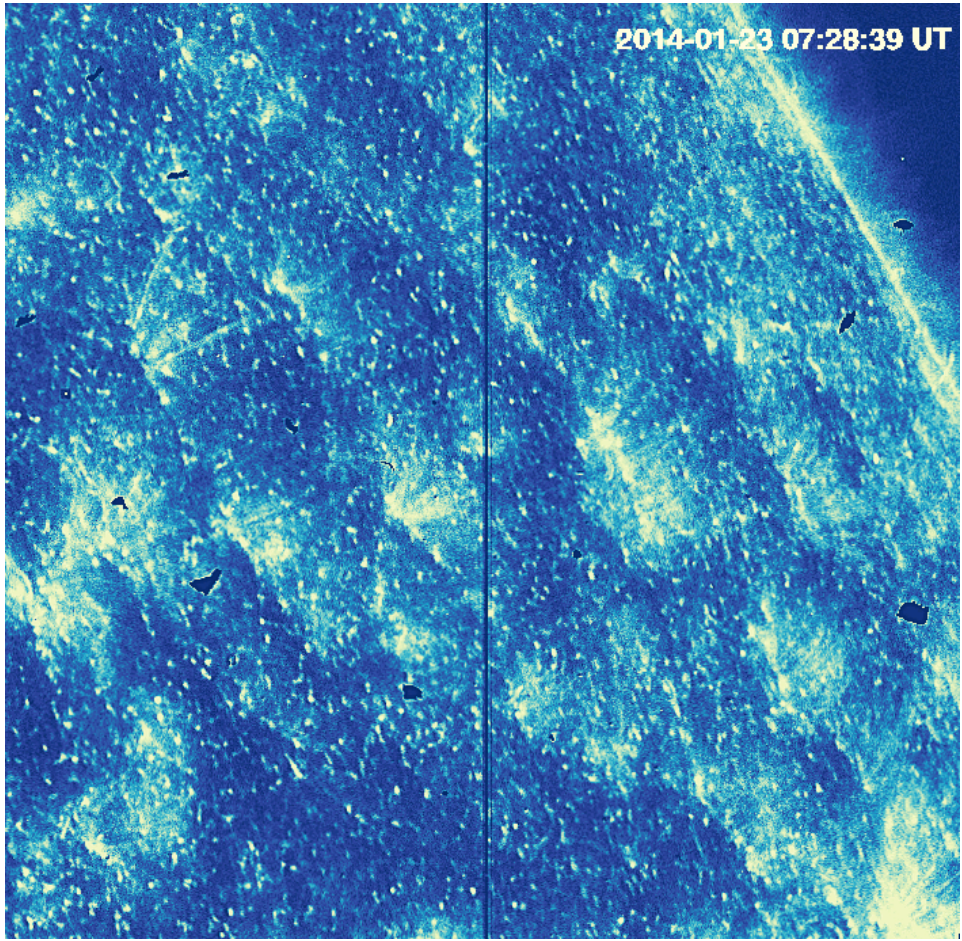

**Fig. S3. Network jets in the 1330Å passband.**
Unsharp masked 1330Å slit-jaw image obtained at 07:28:39 UT on 23 January 2014. The field of view is outlined in Fig. S1B. This is a snapshot of movie S2.

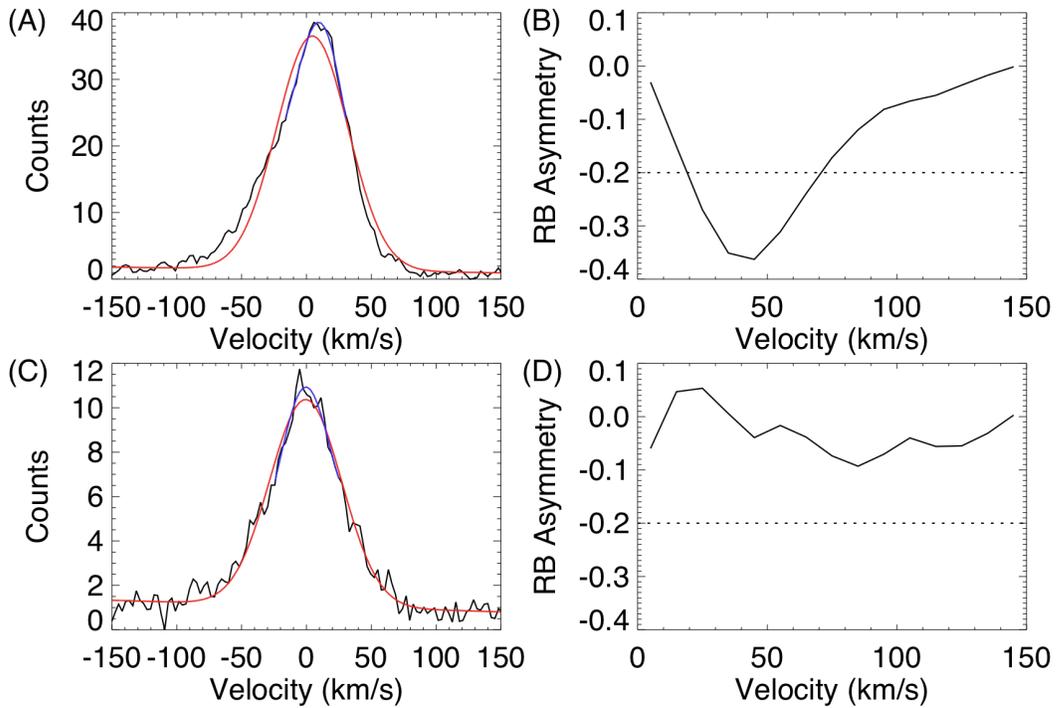

**Fig. S4. Red-blue asymmetry analysis of two line profiles.**
These are results from the 23 January 2014 observation. (A) The line profile at location 1 (Fig.3D). The red and blue lines show the single Gaussian fit results for the full line profile and only the central part of the line profile, respectively. (B) The normalized RB asymmetry profile for this line profile. (C-D) Same as (A-B) but for the profile at location 2 (Fig.3E). The dashed lines in (B) and (D) mark a blueward asymmetry level of 20%.

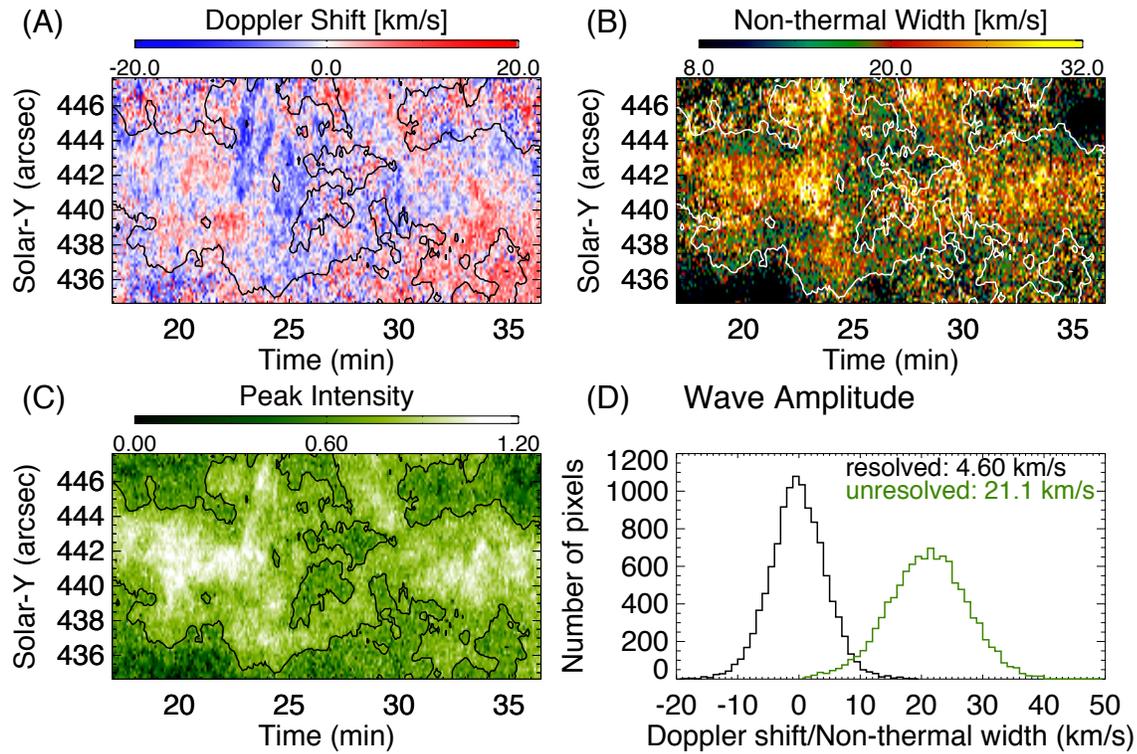

**Fig. S5. Estimation of the resolved and unresolved Alfvén wave amplitudes.**
These are results from the 23 January 2014 observation. (A)-(C) Maps of the line parameters of Si IV 1393.77Å as derived from single Gaussian fit. Pixels with intensity values larger than 0.7 are outlined by the contours and they are regarded as samples of network jets. (D) Distributions of the Doppler shift (black) and non-thermal width (green). The derived average wave amplitudes are shown in the panel.

**Table S1. Summary of the datasets used in this paper.**

| Data sets | Observation time | Pointing (x, y) | Observing mode | Exposure time | FOV of spectral observation | FOV of imaging observation | Cadence of 1330 Å images |
|---|---|---|---|---|---|---|---|
| 1 | 15:56 UT 18 -17:36 UT 18 August 2013 | 784", 368" | Sit and stare | 30s | 0.33"×173" | 173"×173" | 99s |
| 2 | 07:10 UT 23 -08:02 UT 23 January 2014 | 804", 447" | Sit and stare | 4s | 0.33"×119" | 119"×119" | 10s |
| 3 | 23:26 UT 9 - 02:57 UT 10 October 2013 | 513", 284" | Raster scan | 30s | 117"×173" | 283"×173" | 132s |

**Movies**

**Movie S1.** Network jets in different passbands. These are unsharp masked 1400Å, 1330Å, and 2796Å slit-jaw images obtained from 15:56 to 16:46 UT on 18 August 2013.

**Movie S2.** Network jets in the 1330Å passband. These are unsharp masked slit-jaw images obtained from 07:25 to 07:33 UT on 23 January 2014.

**Movie S3.** Examples of network jets. Same as Movie S2 except that only the region of Fig.1A is shown. The spacecraft jitter has been removed through cross-correlation and interpolation of the image sequence.

**Movie S4.** Origin of network jets from small-scale bright regions in the networks. Only the field of view of Fig.2 is shown. The spacecraft jitter has been removed through cross-correlation and interpolation of the image sequence.

**Movie S5.** Signatures of network jets in Si IV 1393.77Å line profiles. (A) Same as Movie S2 except that only a small region around the slit is shown. (B-C) Maps of the peak intensity and line width as derived from a single Gaussian fit to the Si IV 1393.77Å line profiles. The moving vertical line indicates different time.

**Movie S6.** Transition region filamentary structures caused by network jets. These are unsharp masked 1330Å slit-jaw images obtained from 23:26 UT on 9 October 2013 to 02:57 UT on the next day. (A) Movie of 1330Å slit-jaw images. (B-C) Maps of the peak intensity and line width as derived from a single Gaussian fit to the Si IV 1393.77Å line profiles. The moving vertical line indicates the slit location at different time.

These movies are available at

http://www.sciencemag.org/content/346/6207/1255711/suppl/DC1

Additional Movies:

**Movie S7.** Same as Movie S1 but in a small FOV.
http://kurasuta.cfa.harvard.edu/~htian/sciencem2.mov

**Movie S8.** Dynamics of network jets.
http://kurasuta.cfa.harvard.edu/~htian/sciencem1.mov

This paper has been published in Science and the high-resolution version of some figures can be found there:

H. Tian, E. E. DeLuca, S. R. Cranmer, et al., Science 346, 1255711 (2014)